\documentclass[preprint,preprintnumbers,amsmath,amssymb,11pt]{revtex4}
\usepackage{amsmath,epsfig,graphicx,amssymb}

\begin{document}
\title{Skyrme-Random-Phase-Approximation description
of E1 strength in $^{92-100}Mo$}
\author{J. Kvasil, P. Vesely}
\affiliation{Institute of Particle and Nuclear Physics, Charles University,
CZ-18000 Praha, Czech Republic;\\
kvasil@ipnp.troja.mff.cuni.cz}
\author{V.O. Nesterenko, W. Kleinig
\footnote{Permanent address: Tech. Universit\"at Dresden, Inst. f\"ur
Analysis, D-01062, Dresden, Germany.}}
\affiliation{$^{1}$ Bogoliubov Laboratory of Theoretical
Physics, Joint Institute for Nuclear Research, Dubna, Moscow region, 141980,
Russia;\\
nester@theor.jinr.ru}
\author{P.-G. Reinhard}
\affiliation{\it
Institut f\"ur Theoretische Physik II, Universit\"at Erlangen, D-91058,
Erlangen, Germany}
\author{S. Frauendorf
\footnote{Present address:
Department of Physics, University of Notre Dame, Indiana, 46556, USA.}}
\affiliation{
Institut f\"ur Strahlenphysik, Forschungszentrum, Rossendorf, D-01314,
Dresden, Germany}
\date{\today}

\begin{abstract}
The isovector dipole E1 strength in $^{92,94,96,98,100}$Mo is analyzed within
the self-consistent separable random-phase approximation (SRPA) model with
Skyrme forces SkT6, SkM*, SLy6, and SkI3. The special attention is paid to the
low-energy region near the particle thresholds (4-12 MeV), which is important
for understanding of astrophysical processes. We show that, due to a
compensation effect, the influence of nuclear deformation on E1 strength below
10-12 MeV is quite modest. At the same time, in agreement with previous
predictions, the deformation increases the strength at higher energy. At 4-8
MeV the strength is mainly determined by the tail of E1 giant resonance. The
four Skyrme forces differ in description of the whole giant resonance but give
rather similar results below 12 MeV.
\end{abstract}

\maketitle

\section{Introduction}

The isovector  giant dipole resonance (GDR) remains to be a subject of
intense study, now with the accent to exotic nuclei and astrophysical
problems \cite{astrochains}. Besides the GDR is actively used
for inspection and upgrade of the modern self-consistent
mean-field approaches \cite{Ben_rew,Stone_rew}, e.g. those based
on Skyrme forces \cite{Skyrme,Vau}.

In this paper we will investigate GDR in a chain of even-mass isotopes
$^{92-100}$Mo. This chain is of particular interest because here we have
photoabsorption experimental data not only above but also below the particle
emission thresholds, down to 4 MeV \cite{ross_exp_xn,Schwen}. Being rather
weak, the E1 strength near the thresholds is however important for
understanding astrophysical processes, e.g. of the stellar photodisintergation
rate \cite{astro_Arnold,astro_Utsu,astro_Goriely}. In principle, this strength should depend
on such factors as nuclear size and deformation. Since the chain $^{92-100}$Mo
involves spherical (A=92,94,96) and deformed (A=98,100) nuclei
\cite{Mo_def}, it is suitable for estimation of both factors.

The E1 strength in the molybdenum chain has been recently explored in the
random-phase-approximation (RPA) with the phenomenological Nilsson
\cite{Doenau_PRC} and Woods-Saxon \cite{Schwen} single-particle potentials.
In both cases, the important role of the deformation was found. Namely, it was
predicted that prolate/triaxial deformation significantly increases the
E1 strength at 10-14 MeV. However, the calculations \cite{Schwen,Doenau_PRC}
are not self-consistent and so an additional analysis within more involved
microscopic models is desirable. In the present paper we explore the dipole
strength in $^{92-100}$Mo in the framework of the self-consistent separable
RPA (SRPA) with Skyrme forces \cite{nest_PRC_02,p_05,nest_PRC_06}.
SRPA covers both spherical \cite{nest_PRC_02} and deformed
\cite{nest_PRC_06,nest_ijmp_07,nest_ijmp_08} and so is the proper tool for
the present analysis. Factorization of the residual interaction minimizes the
computational effort and therefore allows the systematic exploration. The model was
already used for the analysis of the GDR in different mass regions, including
drip line and superheavy nuclei \cite{nest_ijmp_08,nest_super_08}. Note
that unlike the previous Skyrme-RPA studies of the low-energy E1 strength
\cite{astro_Goriely}, where the GDR deformation splitting was introduced
phenomenologically, SRPA treatment of the deformation effects is fully
self-consistent.

  We will show that conclusions  \cite{Schwen,Doenau_PRC} on the impact of
nuclear deformation should be amended in the sense that the result strictly
depends on the energy region. Namely, the deformation indeed increases the E1
strength at $E >$ 10-12 MeV but, at the same time, has a minor impact at lower
energy, i.e. near the particle thresholds. The later is caused by
a compensation effect of the GDR branches. This effect is quite general
and becomes apparent just in the low-energy regions of astrophysical interest.

\section{Calculation scheme and ground state properties}
\label{sec:calc_scheme}

The calculations are performed within SRPA \cite{nest_PRC_02,p_05,nest_PRC_06}
with the representative set of Skyrme forces, SkT6 \cite{skt6}, SkM* \cite{skms},
SLy6 \cite{sly46}, and SkI3 \cite{ski3}. These forces expose a variety of
features (effective masses, etc) relevant for the GDR
\cite{Ben_rew,Stone_rew,nest_ijmp_08}.
Amongst them, the force SLy6 provides the beest compromise
for the description of the GDR in heavy nuclei
\cite{nest_PRC_06,nest_ijmp_08,nest_super_08}.
The SRPA residual interaction involves contributions
from the time-even densities (nucleon $\rho$, kinetic $\tau$ and spin-orbital
$\vec{\Im}$), time-odd current $\vec j$, direct and exchange Coulomb terms, and
pairing (with delta forces at the BCS level)
\cite{nest_PRC_02,p_05,nest_PRC_06,nest_super_08}.

\begin{figure}[th] \label{fig1}
\centerline{\psfig{file=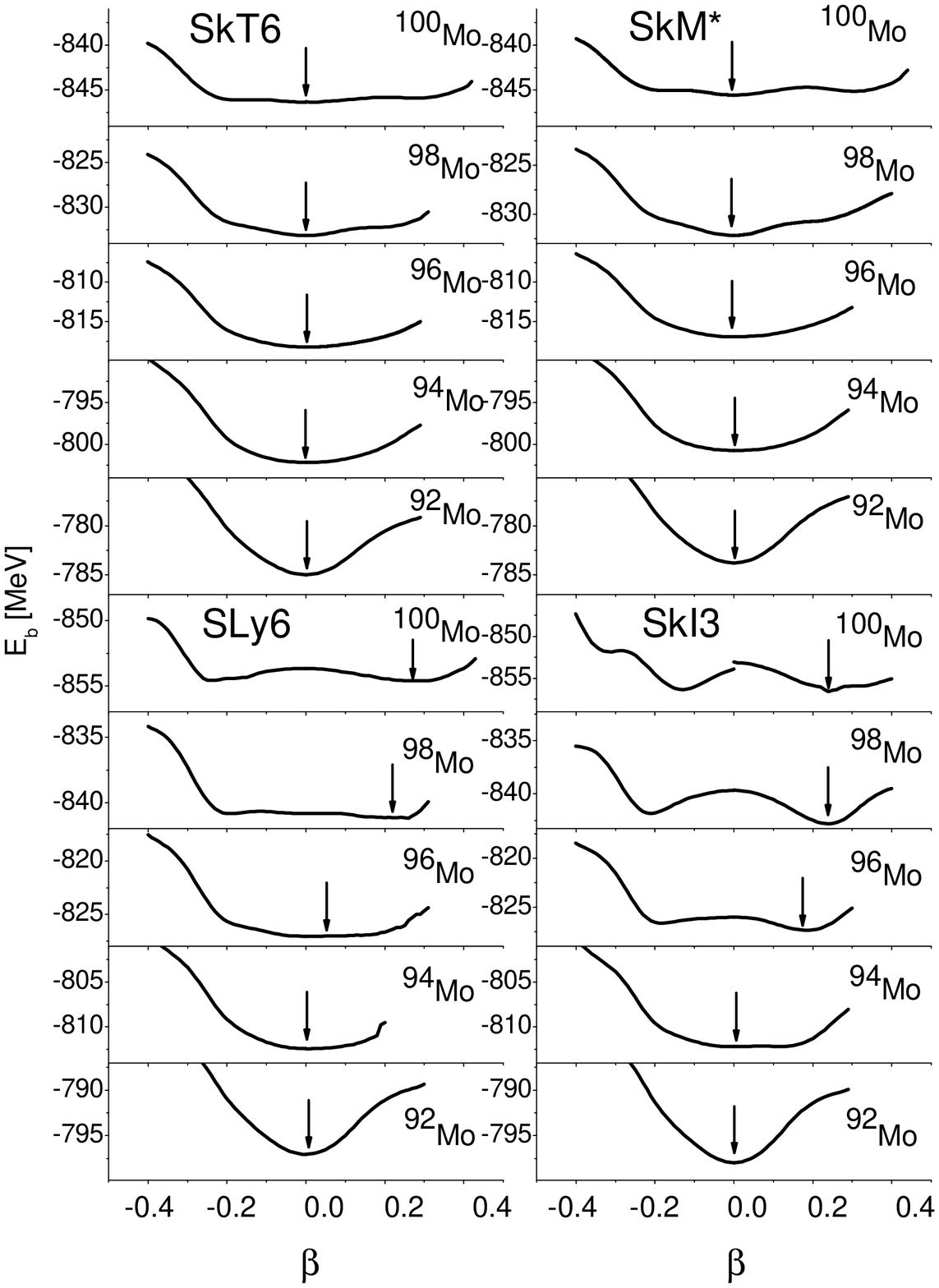,width=9.5cm}} \label{fig:e_bind} \caption{
Energy surfaces (= binding energies) in Mo isotopes, obtained in
quadrupole-constrained mean-field calculations with Skyrme forces SkT6, SkM*,
SLy6, and SkI3. The equilibrium axial deformations $\beta$ are indicated by arrows.}
\end{figure}

The axial equilibrium quadrupole deformation
\begin{equation}\label{eq:quad_def}
\beta = \sqrt{\frac{\pi}{5}} \frac{1}{Z <r^2>_p}\int d\vec r \rho_p(\vec r) r^2 Y_{20}
\end{equation}
(where $\rho_p(\vec r)$ is the proton density in the ground state and
$<r^2>_p=\int d{\vec r}\rho_p(\vec r) r^2/Z$ is the r.m.s. proton radius)
is determined by minimization of the total energy, see Fig. 1 and Table \ref{tab:mo}.
As is seen from Fig. 1, the nuclei $^{94-100}$Mo are soft to $\beta$,
especially the heavy isotopes. Partly this is because of their transition character.
Besides, $^{98}$Mo and $^{100}$Mo are probably
triaxial (with $\epsilon_2=0.18, \gamma=37^{o}$ and $\epsilon_2=0.21, \gamma=32^o$,
respectively \cite{Mo_def,Doenau_PRC}). In the present study the triaxiality is
omitted, which may also lead to a wide plateau and local minima for $\beta < 0$.
Altogether, there is appreciable ambiguity in determination of the
equilibrium deformation $\beta$. In $^{98,100}$Mo we see a significant variation
of $\beta$ with the Skyrme force. While SkT6 and SkM* favor
a spherical shape (in contradicts with significant $\epsilon_2$ in
Refs. \cite{Mo_def,Doenau_PRC}), the forces SLy6 and SkI3
give more reasonable results. So just SLy6 and SkI3 will be mainly used in
the further analysis of the deformation effects. Note, that triaxiality can
cause an additionally spread of the E1 strength. In the present
study  this effect is masked by the proper Lorentz smoothing of the strength.

\begin{table}
\caption{\label{tab:mo} Experimental thresholds for $(\gamma, n)$,
$(\gamma, p)$, $(\gamma, 2n)$, $(\gamma, np)$, and $(\gamma, 2p)$ reactions
\protect\cite{atlas} and equilibrium deformations $\beta$ calculated with
the forces SkT6, SkM*, SLy6, and SkI3 in $^{92-100}$Mo.}
\begin{tabular}{|c|c|c|c|c|c|c|c|c|c|c|}
\hline
A & \multicolumn{5}{|c|}{Thresholds (MeV)}
  & \multicolumn{4}{|c|}{$\beta$}
\\
\cline{2-10}
 & $E_n$ & $E_p$ & $E_{2n}$ & $E_{np}$& $E_{2p}$
  & SkT6 & SLy6 & SkM* & SkI3\\
\hline
 92 & 12.7 & 7.5 & 22.8 & 19.5 & 12.6 & 0.0 & 0.0 & 0.0 & 0.0 \\
 94 & 9.7 & 8.5 & 17.7 & 17.3 & 14.5 & 0.0 & 0.0 & 0.0 & 0.0 \\
 96 & 9.2 & 9.3 & 16.5 & 17.8 & 16.1 & 0.0 & 0.05 & 0.0 & 0.18 \\
 98 & 8.6 & 9.8 & 15.5 & 17.9 & 17.2 & 0.0 & 0.22 & 0.0 & 0.24 \\
100 & 8.3 & 10.1& 14.2 & 18.0 & 19.5 & 0.0 & 0.27 & 0.0 & 0.24 \\
 \hline
\end{tabular}
\end{table}

The photoabsorption (in mb) is computed as \cite{Ring}
\begin{equation}\label{photoab}
 \sigma_{\gamma}(E)= 4.01 \; E \; S(E)
\end{equation}
where
\begin{equation}\label{eq:strength_function}
  S (E) = \sum_{\mu =0,1}
  \sum_{\nu}
  E_{\nu}|\langle\Psi_\nu|\hat{f}_{E1\mu}|\Psi_0\rangle|^2
  \zeta(E - E_{\nu})
\end{equation}
is the strength function with the Lorentz weight
$\zeta =\Delta/(2\pi[(E-E_{\nu})^2+\Delta^2/4]) $
and isovector transition operator
$  \hat{f}_{E1\mu} =
  N/A \sum_{p=1}^Z r_p Y_{1\mu}(\Omega_p)
  -
  Z/A\sum_{n=1}^N r_n Y_{1\mu}(\Omega_n) \;.
$
In Eq. (\ref{eq:strength_function}), $E_{\nu}$ and $\Psi_\nu$ mark eigenvalue
and eigenfunction  of $\nu$-th RPA state, and $\Psi_0$ is the ground state
eigenfunction. The summation runs over the RPA spectrum
until $E_{cut}= 45$ MeV.
The Lorentz function with the averaging parameter
$\Delta$ is used to simulate the broadening effects beyond SRPA
(triaxiality, escape widths, and coupling with complex configurations).
Besides, this smoothing
allows to avoid unnecessary details of the calculated strength which
in any case are not resolved in experiment. Following
\cite{nest_PRC_06,nest_ijmp_08,nest_super_08}, the averaging $\Delta$=2 MeV
is optimal. SRPA allows a direct computation of the
strength function (without finding
the manifold of RPA states), which greatly reduces the effort.

The calculations use a large basis space of single-particle states, from the bottom
of the potential well up to $\sim + 16$ MeV. The integral photoabsorption
\begin{equation}\label{sum_rule}
 \Sigma = \int_{0}^{E_{cut}} dE \; \sigma_{\gamma}(E)
\end{equation}
exhausts  up to 98$\%$ of the energy-weighted sum rule
EWSR $= 9(\hbar e)^2/(8\pi m^*_1)\cdot NZ/A$ with
the isovector effective mass $m^*_1$ arising due to the
velocity-dependent densities $\tau$, $\vec{\Im}$,
and $\vec j$ \cite{nest_ijmp_08,lip_str}. So our basis is
indeed large enough to explore the GDR. The EWSR should not be confused
with the Thomas-Reiche-Kuhn (TRK) sum rule $\Sigma_{TRK} =60 NZ/A$ mb MeV
which deals with the bare nucleon mass $m$.

\section{Results and discussion}
\label{sec:results}

Results of the calculations for E1 strength are given in Figs. 2-7.

In Figs. 2 the photoabsorption in semimagic spherical $^{92}$Mo is presented.
The logarithmic scale is used for a convenient comparison with the
experimental data. Different smoothing parameters $\Delta$ are used. It is seen
that the most appropriate agreement with experiment takes place for $\Delta$=1
and 2 MeV, especially in the low-energy 4-12 MeV of our interest (the same for
other isotopes $^{94-100}$Mo). Taking also into account our previous
results for the GDR \cite{nest_PRC_06,nest_ijmp_08,nest_super_08}, we chose for
the further analysis $\Delta$=2 MeV. As is seen from Fig. 2, for this
averaging we generally reproduce the GDR energy and width. At the same time,
the calculated $\sigma_{\gamma}(E)$ systematically exceeds the experimental
data at the GDR top and falls short at its right flank. This discrepancies can
be partly caused by omitting the complex configurations. Being included, those
configurations could redistribute the strength from the middle to the right
side of the GDR, thus improving agreement with the experiment. Besides,
the discrepancy at the GDR top can be also an artifact of the comparison of
the calculated photoabsorbtion with the separate $(\gamma, p)$ and
$(\gamma, xn)$ channels.

\begin{figure}[t] \label{fig2}
\centerline{\psfig{file=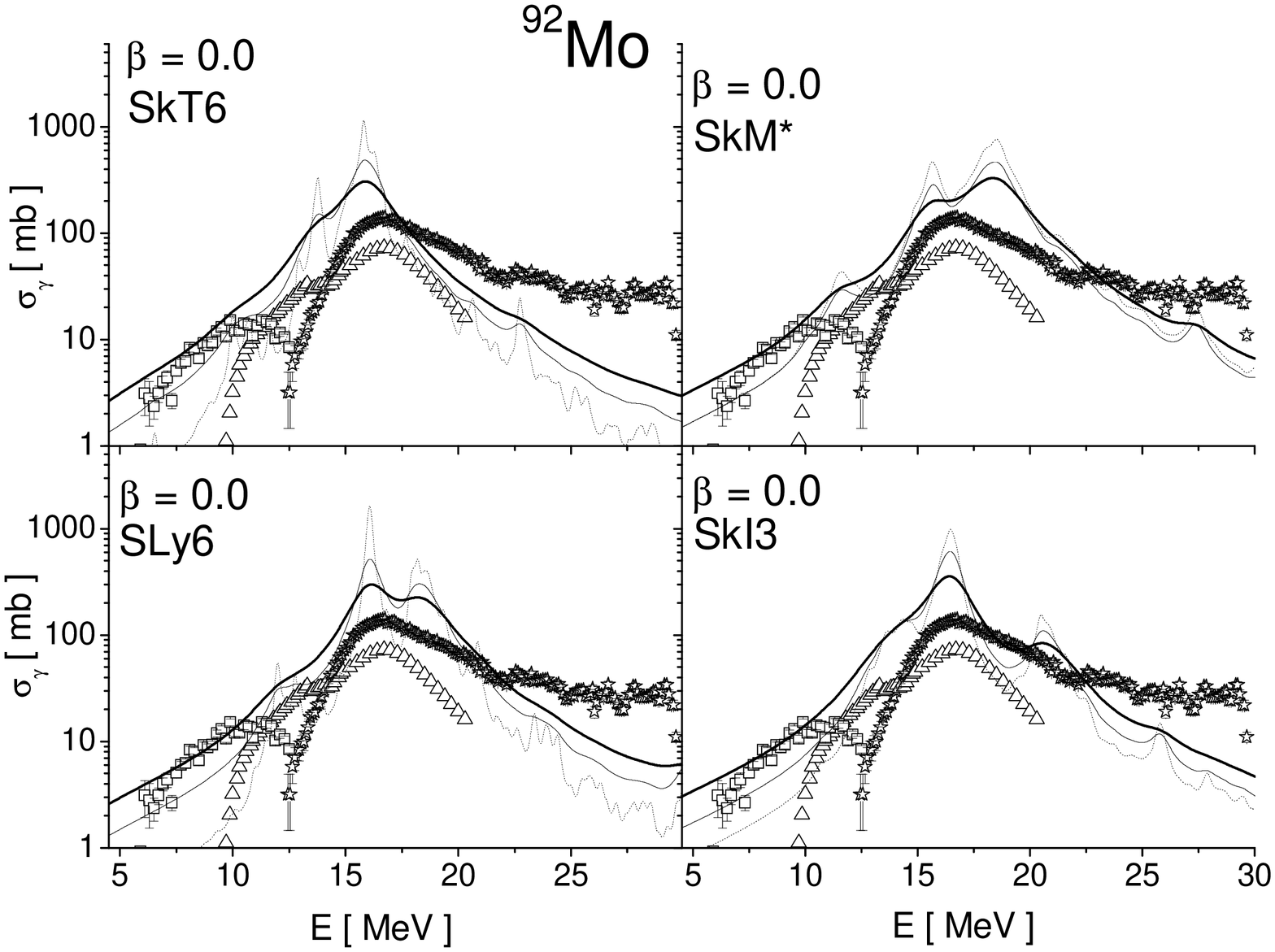,width=7.0cm,angle=-90}}
\caption{Photoabsorption (in logarithmic scale) in $^{92}$Mo
calculated with Skyrme forces
SkT6, SkM*, SLy6, and SkI3 with the averaging $\Delta$=0.25 MeV (dotted curve),
1 MeV (solid curve) and 2 MeV (bold solid curve). The experimental
$(\gamma, \gamma')$, $(\gamma, p)$, and $(\gamma, xn)$
data \protect\cite{ross_exp_xn,Schwen} are given by boxes, triangles, and stars
in the energy intervals 5.1-12.5, 7.9-20.3, and 12.4-29.7 MeV,
respectively.
}
\end{figure}

\begin{figure}[th] \label{fig3}
\centerline{\psfig{file=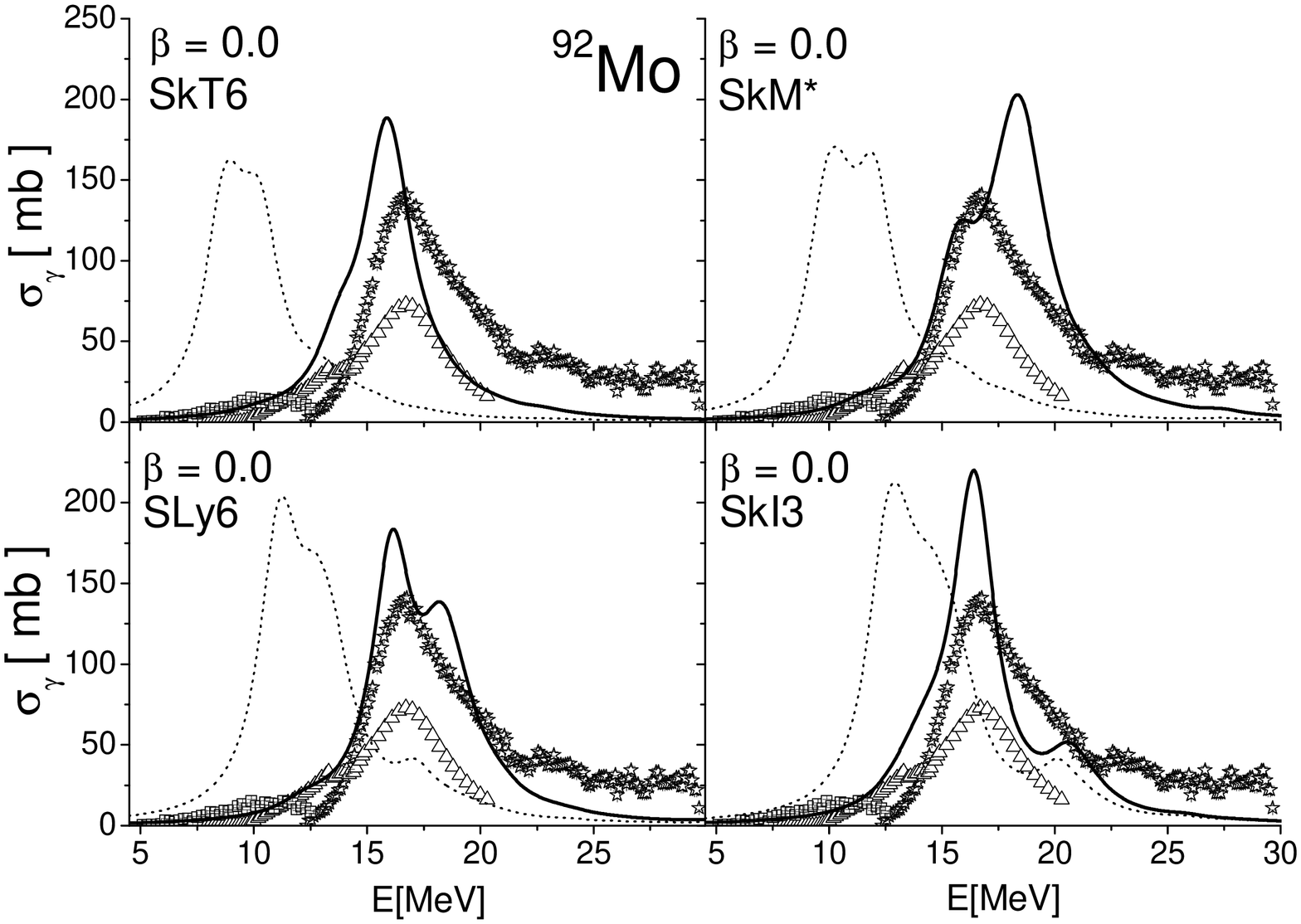,width=7.0cm,angle=-90}}
\caption{The same as in Fig. 2 but for linear scale and averaging $\Delta$=2
MeV only. The dash curve exhibits the unperturbed two-quasiparticle (2qp)
results.
}
\end{figure}
\begin{figure} \label{fig4}
\centerline{\psfig{file=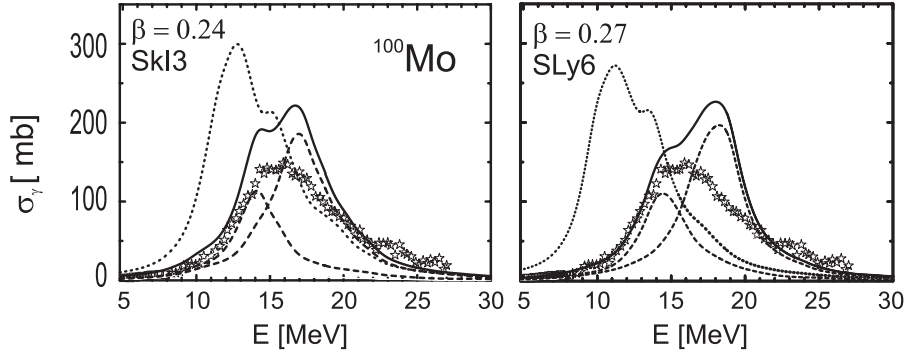,width=12.0cm}}
\caption{The same as in Fig. 3 for $^{100}$Mo and forces SLy6 and SkI3.
The experimental data \protect\cite{ross_exp_xn,Schwen} for $(\gamma, \gamma')$
at 4.1-8.1 MeV and $(\gamma, xn)$ at 8.3-27 MeV
are given by boxes and stars, respectively.
The GDR branches $\mu=0$ (small bump) and $\mu=1$ (twice larger bump)
exhibited by the dash curve demonstrate the deformation
splitting of the resonance.
}
\end{figure}

In the logarithmic scale the photoabsorption obtained for different Skyrme
forces looks rather similar. To distinguish the difference between predictions
of different forces, one should switch to the linear scale, which is done in
Figs. 3 and 4 for semimagic $^{92}$Mo and deformed $^{100}$Mo.
Now it is seen that in spherical $^{92}$Mo the best description is
for SLy6. For deformed $^{100}$Mo only SLy6 and SkI3 results are
presented. The forces SkT6 and SkM* predict for $^{100}$Mo
the spherical shape and so obviously fail here.
The calculations for $^{100}$Mo overestimate the
experimental data (as photoabsorption over $(\gamma, xn)$) but
keep EWSR and provide an
acceptable agreement for the GDR energy centroid and width.

As was mentioned in the introduction, the E1 strength near the particle
thresholds  is of particular interest for some
astrophysical problems \cite{astro_Arnold,astro_Utsu}. In this connection,
it is important to understand the main physical mechanisms responsible for the
evolution of this strength with the neutron number N in the isotope
chain. First of all, this evolution is determined by the empirical rule
$E_{GDR}=81 A^{-1/3}$ MeV \cite{Ring} relating the GDR energy and
nuclear mass number A. The higher N (and so A), the more the
GDR downshift and stronger the GDR tail (and relevant E1 strength)
near the thresholds. This size factor seems to dominate.
However, two other effects, nuclear deformation and internal E1 strength
in the region (pygmy resonance), can also come to play and influence
the general size trend.

These effects are inspected in Figs. 5-7. The impact of deformation is
illustrated in Fig. 5, where the E1 strength calculated at
zero and non-zero deformations is compared with the experimental data
\cite{ross_exp_xn,Schwen}. We get the nonzero deformation only in
$^{98}$Mo and $^{100}$Mo for the forces SLy6 and SkI3. In all other cases the
calculations give $\beta=$0, see Table 1.
Fig. 5 shows that for $\beta$=0 all the forces provide more or less acceptable
agreement with the experiment, the best for SkT6 and SkI3. Inclusion of the
deformation changes the results. While at $E>12$ MeV we have, in accordance with
Refs. \cite{Schwen,Doenau_PRC}, increasing E1 strength, in the interval of
our main interest, $E<12$ MeV, we see its modest decrease.
The effect is prominent and even stronger than the change in E1 strength for
the neighbor isotopes.

\begin{figure} \label{fig5}
\centerline{\psfig{file=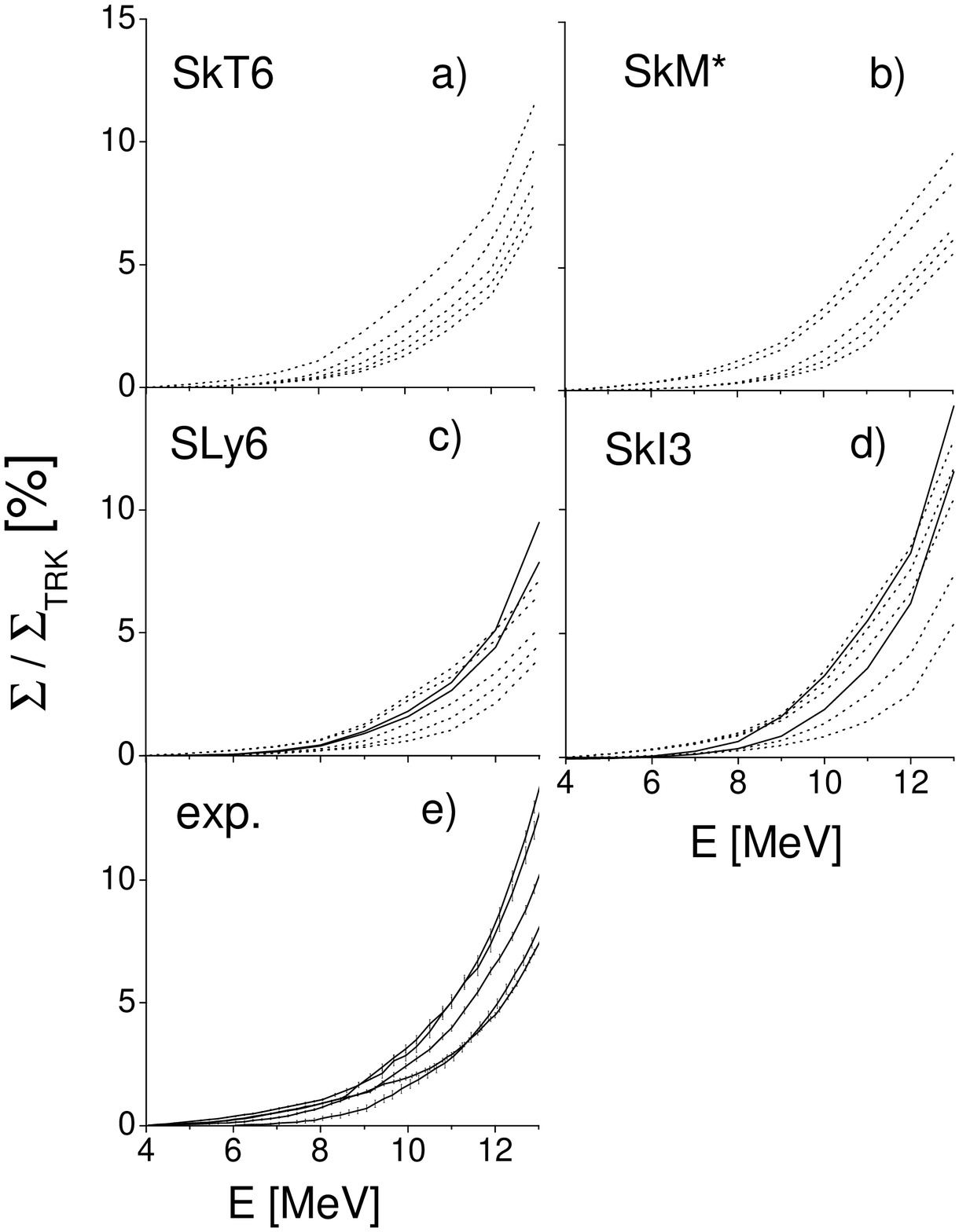,height=10.0cm,width=9.0cm}}
\caption{The calculated a)-d) and experimental
\protect\cite{ross_exp_xn,Schwen} e) integral
low-energy photoabsorption in Mo isotopes. The calculated
results are exhibited for non-zero (= equilibrium
in $^{98,100}$Mo for SLy6 and SkI3) and zero (= equilibrium in other cases)
deformations by solid and dotted curves, respectively.
In all the panels the sequence of
curves for $^{92,94,96,98,100}$Mo has the same order: from the lowest
for $^{92}$Mo to the highest  for $^{100}$Mo.
}
\end{figure}

The deformation effect for $E<12$ MeV can be explained (for both
prolate and oblate shapes) by destructive competition of two factors, deformation
shifts of $\mu=0$ and $\mu=1$ branches of the GDR. Indeed, the deformation shifts
the GDR branches in opposite directions thus minimizing the resulting deformation
impact. Though the branch $\mu=$1 is twice stronger than $\mu=$0 one, its deformation
shift is less, hence a strong mutual compensation of both $\mu=0$ and  $\mu=1$
deformation impacts. Being strong, the compensation is usually not complete. So,
depending on the concrete case, one may finally observe a modest decrease or increase of
the low-energy E1 strength. Of course, the compensation holds at the energies
far enough from the lower GDR branch (which is just the case for $E<12$ MeV).
Instead, while approaching the lower GDR branch, we always gain E1 strength
with the deformation, as was earlier found in Refs. \cite{Schwen,Doenau_PRC}.

These arguments are illustrated in Fig. 6 for prolate $^{100}$Mo.
The calculations were done for SLy6 with averaging parameters
$\Delta$ = 2 and 1 MeV.
It is seen that for $\Delta$ = 2 the deformation-induced
$\mu=0$ and 1 contributions indeed strictly compensate each
other, thus leading to a slight decrease of E1 strength at $9<E<12$ MeV
and no effect at the lower energy. Instead, in accordance with
\cite{Schwen,Doenau_PRC}, the strength grows at $E>12$ MeV.
The minor deformation effect at $9<E<12$ MeV becomes invisible for the less
averaging $\Delta$ = 1 MeV, which is explained by weakening the GDR low-energy tail.
Note that for $\Delta$ = 2 MeV we get much better agreement with the experiment
than for $\Delta$ = 1 MeV, which additionally justifies the large averaging as
the best choice.

\begin{figure} \label{fig6}
\centerline{\psfig{file=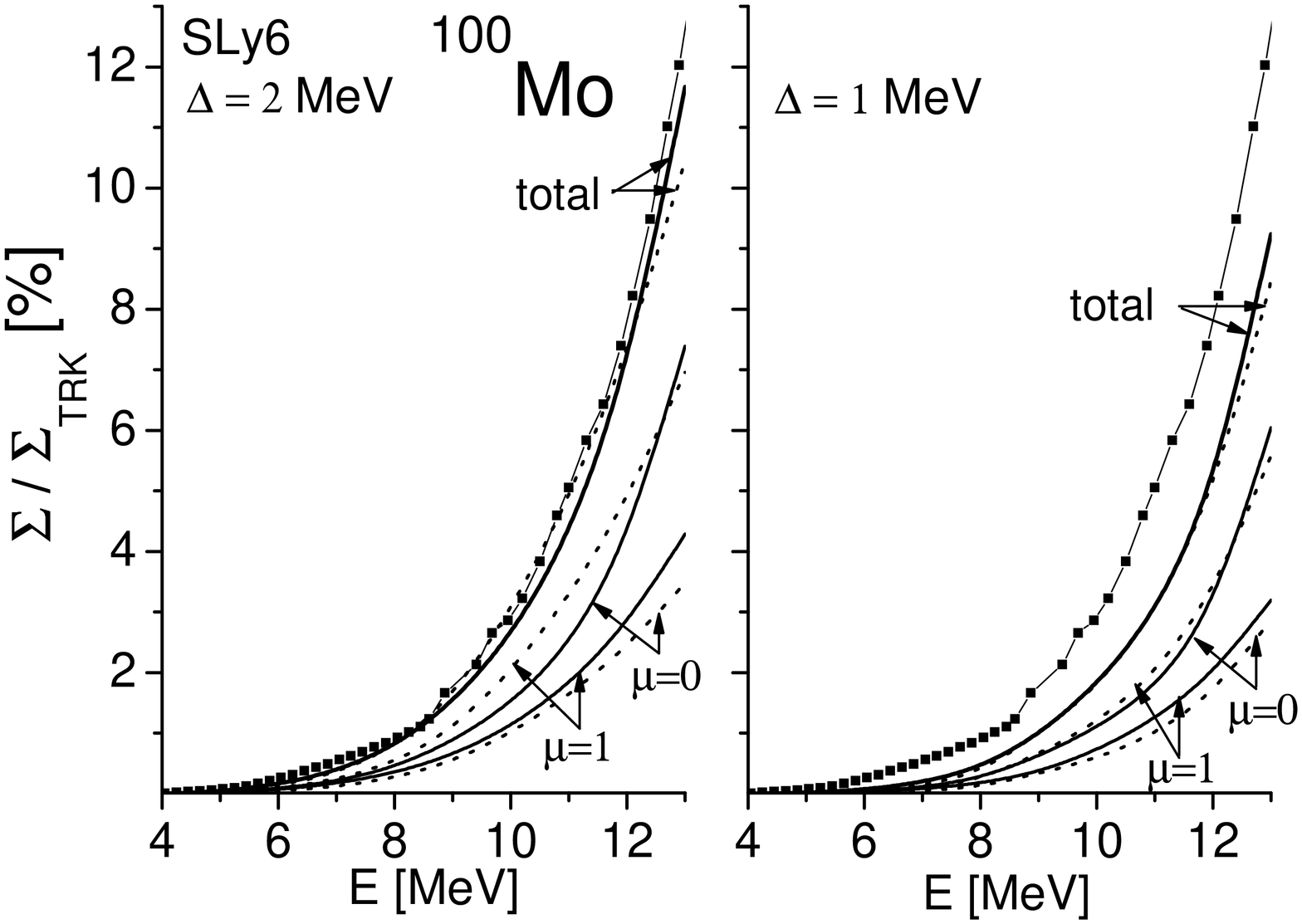,width=5.0cm,angle=-90}}
\caption{The low-energy photoabsorption in $^{100}$Mo,  calculated
with the force SLy6 for $\beta=$0 (dotted curves) and
equilibrium deformation $\beta=$0.27 (solid curves). The averaging
$\Delta=$2 MeV (left) and 1 MeV (right) is used. The total strength
as well as the strengths of the branches $\mu=0$ and $\mu=1$
are indicated by arrows. The experimental data
\protect\cite{ross_exp_xn,Schwen} are depicted by full boxes.
}
\end{figure}
\begin{figure} \label{fig7}
\centerline{\psfig{file=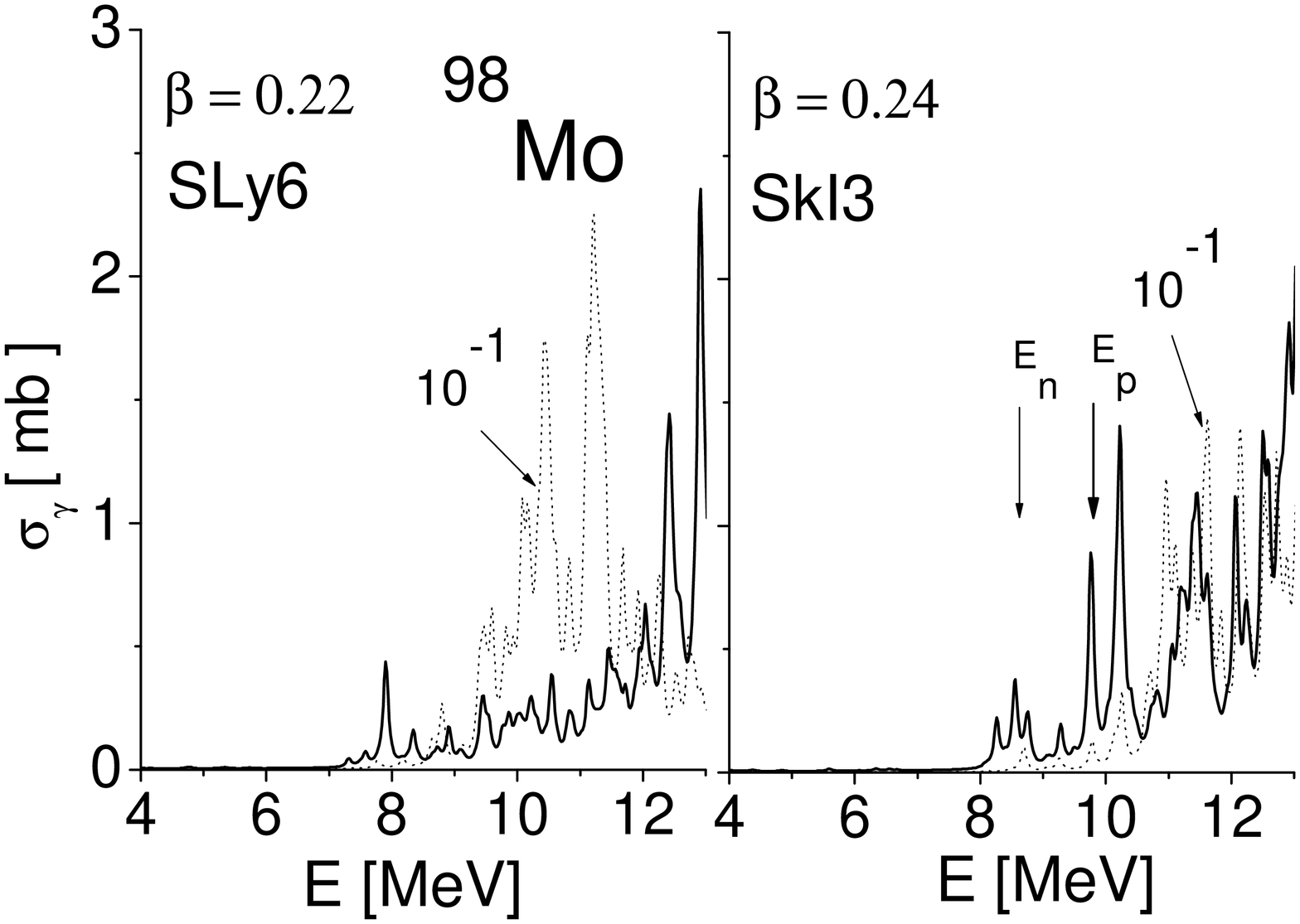,height=7cm,width=5cm,angle=-90}}
\caption{The low-energy photoabsorption in deformed $^{98}$Mo,
calculated for the forces SLy6 and SkI3 with (solid curve)
and without (dash curve) the residual interaction. To see the fine
structure, the small averaging $\Delta=$0.1 MeV is used.
The unperturbed photoabsorption is decreased 10 times as indicated. The vertical
arrows show ($\gamma, n$) and ($\gamma, p$) thresholds.
}
\end{figure}

Finally, Fig. 7 demonstrates for deformed $^{98}$Mo the role
of internal E1 excitations in forming the low-energy strength. To make visible
a fine structure,  the small averaging $\Delta=$0.1 MeV is used. It is seen
that for both forces the E1 strength between 4 and 7-8 MeV does not
exhibit any internal structure while the structure at 7-9 MeV is weak.
So, the E1 strength for $E<9$ MeV is mainly determined by the GDR tail
(the same for other Mo isotopes).
This means that all the effects discussed above for the GDR are indeed
relevant for the energies near and below the particle thresholds.
The absence of E1 structures at 4-8 MeV is natural since 2qp
excitations with $\Delta\cal{N}$=1 ($\cal{N}$ is a principle shell
quantum number) lie at a higher energy, see Figs. 3 and 4.
Note that  for $E>9-10$ MeV the E1 strength rises and exhibits more
appreciable structure (perhaps the pygmy resonance observed in
\cite{ross_exp_xn}).

\section{Summary}
\label{sec:summary}

The E1 strength in $^{92,94,96,98,100}$Mo is investigated in the framework of
the self-consistent separable RPA method for the set of Skyrme forces SkT6, SkM*, SLy6,
and SkI3. The main attention is paid to low-energy strength below the particle
thresholds, which is of a keen interest for astrophysical problems
\cite{astro_Arnold,astro_Utsu}. To our knowledge, this is the first Skyrme-RPA study
of E1 strength in Mo isotopes with the self-consistent treatment
of deformation effects. Some important factors (triaxiality, coupling with
complex configurations, escape widths) are simulated by the Lorentz smoothing
and others (pairing impact, energy dependence of the smoothing) need an
additional analysis, which makes our description of E1 strength yet tentative.
Nevertheless, some useful conclusions can be done.

We confirmed our previous statement
\cite{nest_PRC_06,nest_ijmp_07,nest_ijmp_08,nest_super_08} that
the force SLy6 with the Lorentz smoothing $\Delta =2$ MeV  gives
the most reasonable description of the whole GDR. The low-energy E1 strength
is shown to be mainly determined by the GDR tail. In spherical $^{92,94,96}$Mo,
this strength is well described by all the forces.

It is found that the deformation impact in the low-energy E1 strength
depends on the particular energy interval. While approaching the GDR,
$E>12$ MeV, we get a definite increment of the strength with the deformation
(as was earlier found in Refs. \cite{Schwen,Doenau_PRC}).
However, at $E<12$ MeV, i.e. near and below the particle thresholds,
the deformation impact almost vanishes because of the compensation
of the deformation contributions from $\mu=0$ and $\mu=1$ GDR branches.
The effect of triaxiality should still be checked.
We plan to extend our exploration
to Nd and Sm isotope chains where the triaxiality is absent.

\section*{Acknowledgments}
We thank Prof. F. D\"onau for initiation of this work and fruitful
discussions.
The work was partly supported  by the DFG grant RE 322/11-1 and
grants Heisenberg - Landau (Germany - BLTP JINR) and Votruba - Blokhintsev
(Czech Republic - BLTP JINR) for 2008 year.
W.K. and P.-G.R. are grateful for the BMBF support under contracts
06 DD 139D and 06 ER 808. This work is also a part of the research
plan MSM 0021620859 supported by the Ministry of Education of the
Czech Republic. It was partly funded by Czech grant agency
(grant No. 202/06/0363) and grant agency of
Charles University in Prague (grant No. 222/2006/B-FYZ/MFF).

\end{document}